\documentclass{pasj00}
\draft

\begin{document}
\SetRunningHead{Nakanishi et al.}{Dense and Warm Molecular Gas in NGC 6240}
\Received{2004 December 13}
\Accepted{2005 June 6}


\title{Dense and Warm Molecular Gas between Double Nuclei of the Luminous Infrared Galaxy NGC 6240}

\author{Kouichiro \textsc{Nakanishi}, and Sachiko K. \textsc{Okumura}}
\affil{Nobeyama Radio Observatory, Minamimaki, Minamisaku, Nagano 384-1305}\email{nakanisi@nro.nao.ac.jp, sokumura@nro.nao.ac.jp}

\author{Kotaro \textsc{Kohno}}
\affil{Institute of Astronomy, The University of Tokyo, 2-21-1 Osawa, Mitaka, Tokyo 181-0015}\email{kkohno@ioa.s.u-tokyo.ac.jp}

\author{Ryohei \textsc{Kawabe}}
\affil{National Astronomical Observatory of Japan, 2-21-1 Osawa, Mitaka, Tokyo 181-8588}\email{kawabe@nro.nao.ac.jp}

\and

\author{Takao \textsc{Nakagawa}}
\affil{Institute of Space and Astronautical Science,
Japan Aerospace Exploration Agency (JAXA),\\
3-1-1 Yoshino-dai, Sagamihara, Kanagawa 229-8510}
\email{nakagawa@ir.isas.jaxa.jp}


\maketitle


\begin{abstract}
High spatial resolution observations of 
the $^{12}$CO(1--0), HCN(1--0), HCO$^+$(1--0), and $^{13}$CO(1--0) 
molecular lines toward the luminous infrared merger NGC 6240 
have been performed using the Nobeyama Millimeter Array 
and the RAINBOW Interferometer.
All of the observed molecular emission lines
are concentrated in the region between the double nuclei of the galaxy. 
However, the distributions of both HCN and HCO$^+$ emissions
are more compact compared with that of $^{12}$CO, 
and they are not coincident with the star-forming regions.
The HCN/$^{12}$CO line intensity ratio is 0.25; this suggests that 
most of the molecular gas  between the double nuclei is dense.
A comparison of the observed high HCN/$^{13}$CO intensity ratio, 5.9, with
large velocity gradient calculations suggests that the molecular gas is
dense [$n({\rm H_2})=10^{4-6}$ cm$^{-3}$] and warm ($T_{\rm kin}>$50 K).
The observed structure in NGC 6240 may be explained by
time evolution of the molecular gas and star formation,
which was induced by an almost head-on collision or 
very close encounter of the two galactic nuclei
accompanied with the dense gas and star-forming regions.
\end{abstract}


\section{Introduction}

Luminous infrared galaxies (LIRGs) and ultra-luminous infrared galaxies 
(ULIRGs) are known to be huge molecular gas reservoirs 
[$M({\rm H}_2)\gtrsim10^9$ \MO].
The results of $^{12}$CO(1--0) and (2--1) observations reveal that
this molecular gas is usually concentrated at the center of the galaxies
(e.g. \cite{Okumura1991, Bryant1999}),
and in general it forms a compact (radius $\lesssim1$ kpc) 
rotating disk or ring (e.g., \cite{Downes1998, Sakamoto1999}).
The molecular gas within the disk or ring often shows large turbulent motions,
and the molecular line profiles exhibit large velocity widths of
up to 1000 km s$^{-1}$.
The physical properties, namely the density and the temperature, 
of the molecular gas in LIRGs/ULIRGs are still being studied.
Single-dish HCN(1--0) observations (\cite{Solomon1992};
\cite{Gao2004a},\yearcite{Gao2004b}) revealed that most of their molecular gas
is supposed to be dense [$n({\rm H}_2)>10^4$ cm $^{-3}$].
Single-dish observations of higher excitation $^{12}$CO and $^{13}$CO 
emissions suggest there is large amount of warm molecular gas 
(\cite{Aalto1995,Mauersberger1999,Yao2003}). 
For the denser and warmer molecular gas tracers (e.g. HCN, $^{13}$CO), however,
the spatial distributions in LIRGs/ULIRGs are still unknown,
although the physical properties of gas are keys to understand the 
activity of LIRGs/ULIRGs (starburst in the nuclear region and/or AGN).
Moreover, the spatial variations of the molecular gas density and the temperature
in the circumnuclear regions of LIRGs/ULIRGs are still unclear.
For such studies, high angular resolution observations of molecular lines
with radio interferometers are needed, but 
only a few observations have been performed 
so far (e.g. \cite{Aalto1997,Radford1991}).

NGC 6240 ($\alpha=$\timeform{16h50m27s.83},
$\delta=$\timeform{+2D28'58''.1} (B1950);
$D=98$ Mpc, if we adopt $H_0= 75$ km s$^{-1}$ Mpc$^{-1}$; 
\timeform{1''}=475 pc) is one of the most well-known nearby LIRGs,
with an infrared luminosity of $L_{\rm IR}\sim 6\times10^{11}\LO$
\citep{Sanders1996}.
This galaxy is a merger, with clearly visible disturbed morphology 
and tidal tails in optical images.
NGC 6240 also has double nuclei detected in a wide range of wavelengths
from X-rays to radio
(\cite{Komossa2003, Thronson1990, Scoville2000, Tecza2000, Colbert1994}).
The apparent separation of the nuclei is \timeform{1.''5} (710 pc)
in the 15 GHz radio continuum map \citep{Carral1990}.
Each of the nuclei are accompanied by an intense massive star-forming 
region (e.g. \cite{AlonsoHerrero2002}),
and both of them also host an AGN (\cite{Komossa2003}).
Interferometric $^{12}$CO(1--0) and (2--1) observations have been
carried out by many groups (\cite{Okumura1991, Bryant1999, Tacconi1999}).
These observations show that about half of the molecular gas is 
concentrated in a rotating thick disk-like structure 
($\sim 500$ pc diameter) between the double nuclei,
but the motion of the gas is highly disturbed and 
shows large velocity dispersions.
The total mass of the molecular gas within a radius of $\sim500$ pc 
is estimated to be 2--4$\times10^{9} \MO$, which constitutes about half of
the dynamical mass \citep{Tacconi1999}.
Intense HCN(1--0) emission has been detected with the IRAM 30 m telescope
(\cite{Solomon1992}). 
The obtained HCN/$^{12}$CO(1--0) intensity ratio is 0.15,
which is a typical value for starburst galaxies, 
and suggests the presence of large amounts of dense molecular gas.

In this paper, we present the results of 
interferometric observations of molecular emission lines toward 
the central region of NGC 6240. 
The main goal of this study is to reveal the location and distribution 
of dense gas and to constrain the physical properties of the molecular gas.
We describe the observational parameters and data-reduction procedures
in section 2, and the observational results of the
$^{12}$CO, HCN, HCO$^+$, and $^{13}$CO emission lines and continuum emission
in section 3.
In this section, we also refer to the observed molecular line intensity
ratios and those characteristics.
With these data, the physical properties of the molecular gas 
are discussed in section 4, and the evolutionary scenario of molecular gas
and star formation according to the merging process is described in section 5. 
Finally, we summarize our study in section 6.


\section{Observations and Data Reduction}

Aperture synthesis observations toward NGC 6240 were carried out
with the Nobeyama Millimeter Array (NMA) and the RAINBOW Interferometer
at the Nobeyama Radio Observatory (NRO) between 1995 November -- 1996 March,
and 2000 January -- 2003 April.
The NMA consists of six 10 m-antennas equipped with cooled DSB SIS receivers.
The RAINBOW Interferometer is a 7-element array combining the NMA 
with the NRO 45 m telescope,
providing higher spatial resolution and sensitivity than those of the NMA.

We observed NGC 6240 in the following emission lines:
$^{12}$CO(1--0), HCN(1--0), HCO$^+$(1--0), and $^{13}$CO(1--0).
Except for the $^{12}$CO observations, the backend we used was 
the Ultra-Wide-Band Correlator (UWBC; \cite{Okumura2000}).
For the $^{12}$CO observations, the old-FX correlator was employed
(\cite{Chikada1987}).
The quasars B1548+015, B1656+053, and B1655+077 were used for phase and
amplitude reference calibrations;
3C 279, 3C 345, and 3C 454.3 were used for bandpass calibration.
Uranus, Neptune, and Mars were used as absolute flux-scale calibrations
for the amplitude calibrator.
The uncertainties in the absolute flux scale were estimated to be $\sim10$\%
for each observation.
The HCN and HCO$^+$ emissions were observed simultaneously within the same
correlator passband.
The uv-data were calibrated with the UVPROC-II software package developed at
NRO (\cite{Tsutsumi1997}), and then imaged with natural
UV weighting, and CLEANed, with the NRAO AIPS package.

The detailed observational parameters are summarized in table \ref{obslog}.


\section{Results}

In this section we describe the distributions and kinematical properties 
of the observed emission lines and calculate the intensity ratios.
Our HCO$^+$ and $^{13}$CO interferometric observations are 
the first ones on NGC 6240.
The observational results of continuum emission at 87 GHz 
and 108 GHz are also presented.
Table \ref{obsres} summarizes the observational results.

\subsection{Properties of Emission Lines}
 
\begin{figure}
\begin{center}
\FigureFile(160mm,220mm){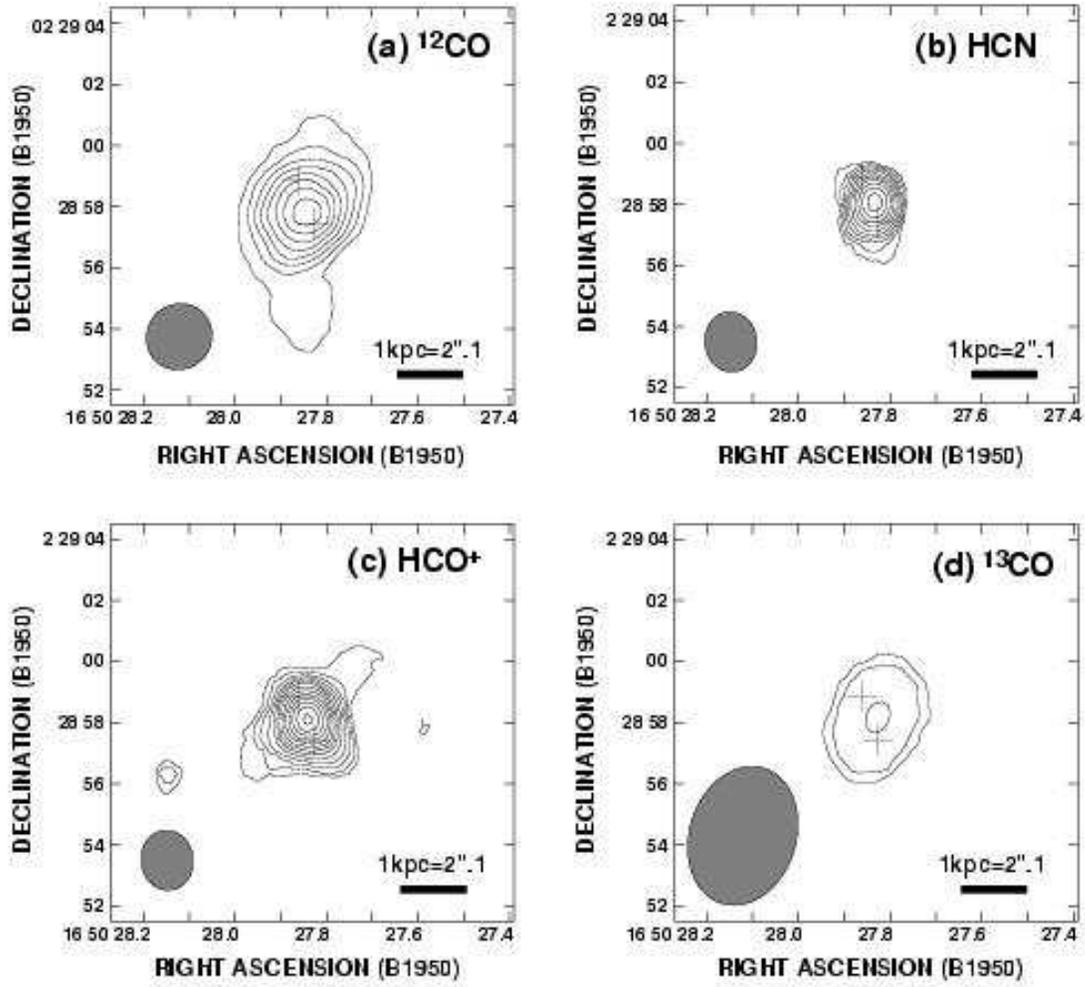}
\end{center}
\caption{
Integrated intensity maps of (a) $^{12}$CO(1--0), (b) HCN(1--0),
(c) HCO$^+$(1--0), and (d) $^{13}$CO(1--0) emission in NGC 6240.
The contour levels in each map are
(a) 2, 4, 6, ..., 16$\sigma$, where $1\sigma = $ 2.6 Jy km s$^{-1}$, 
(b) 2, 3, 4, ..., 11$\sigma$, where $1\sigma = $ 0.79 Jy km s$^{-1}$,
(c) 2, 3, 4, ..., 14$\sigma$, where $1\sigma = $ 0.78 Jy km s$^{-1}$, and
(d) 2, 3, 4$\sigma$, where $1\sigma = $ 0.94 Jy km s$^{-1}$.
The synthesized beam is shown at the bottom-left corner of each map.
The crosses show the peak positions of 15 GHz continuum (\cite{Carral1990}),
which coincide with the positions of the double nuclei (AGNs).
No primary beam correction has been applied.
}\label{intmaps}
\end{figure}

\begin{figure}
\begin{center}
\FigureFile(160mm,220mm){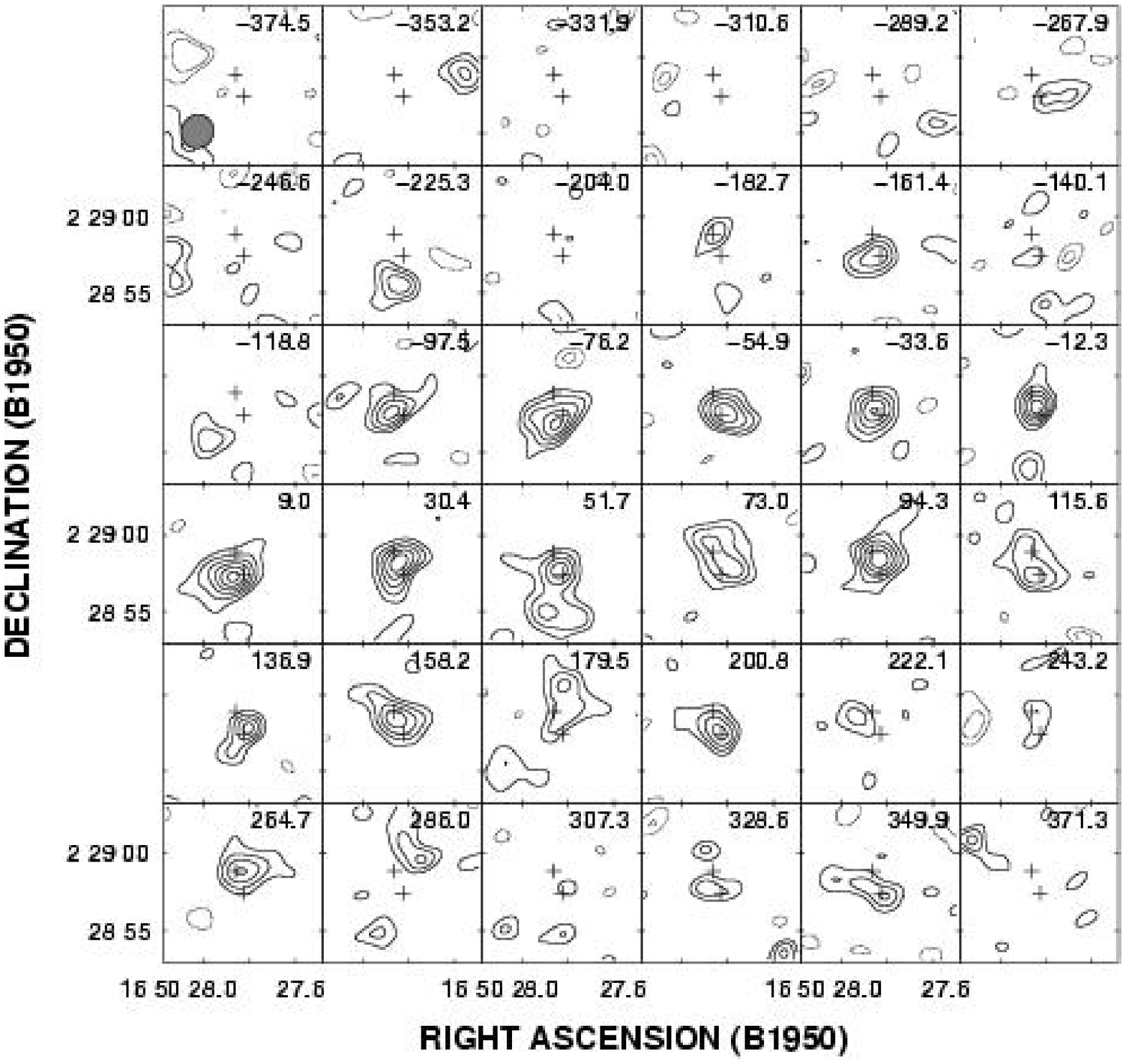}
\end{center}
\caption{
Velocity channel maps of the $^{12}$CO(1--0) emission in the central 
\timeform{10''}$\times$\timeform{10''} 
(4.8 kpc $\times$ 4.8 kpc) region of NGC 6240.
The contour levels are $-3$, $-2$, 2, 3, 4, 5, 6$\sigma$, where
$1\sigma =$ 20 mJy beam$^{-1}$ (negative contours are dashed).
The labels in each channel (top right) are velocity offsets, 
in km s$^{-1}$, from the systemic velocity ($v_{\rm LSR}$) of 7339 km s$^{-1}$
\citep{Downes1993}.
The velocity width of each channel is 21.3 km s$^{-1}$.
The crosses show the peak positions of the 15 GHz continuum emissions
(\cite{Carral1990}).
}
\label{cochmap}
\end{figure}

\begin{figure}
\begin{center}
\FigureFile(80mm,220mm){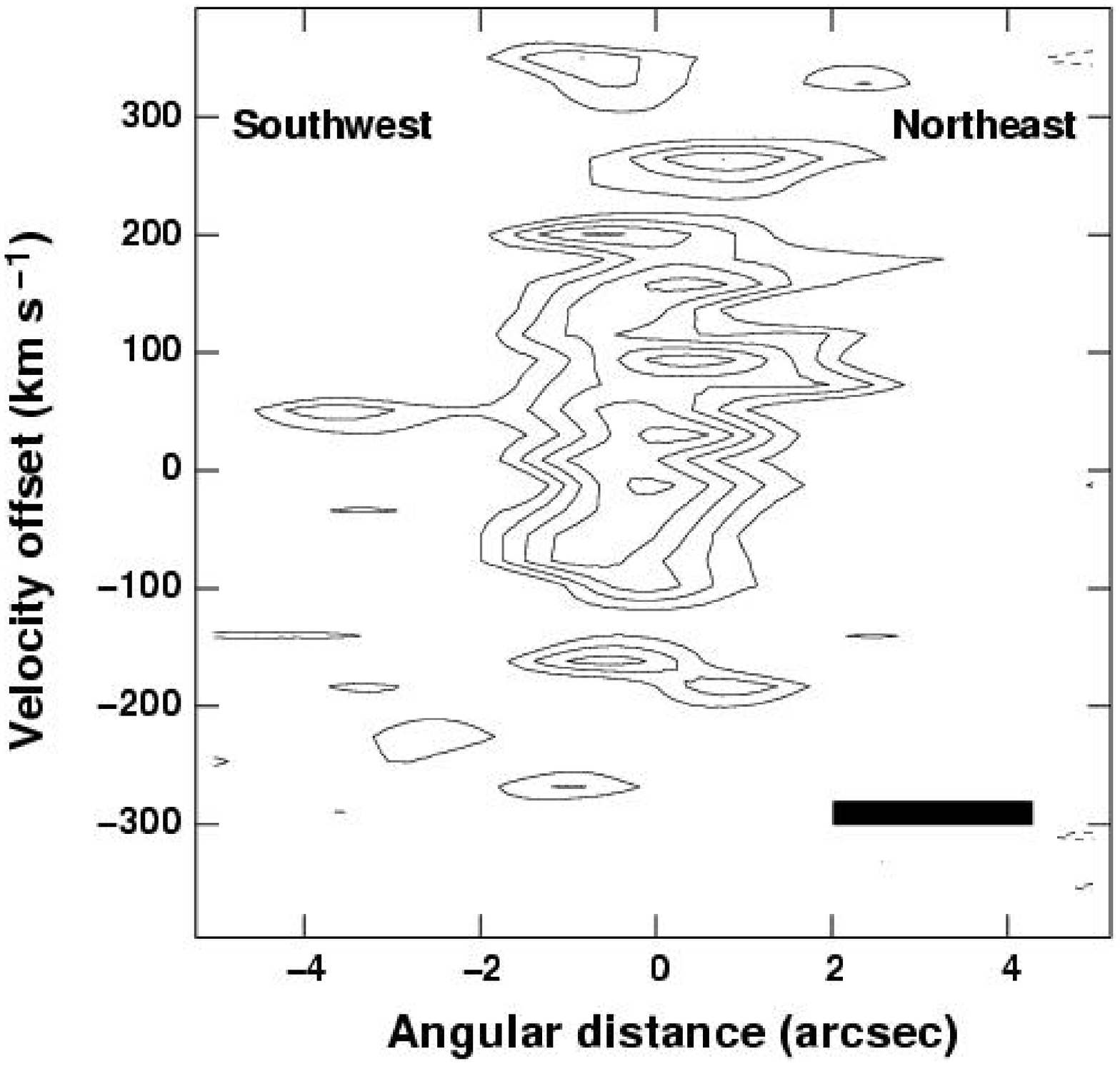}
\end{center}
\caption{
Position--velocity map of $^{12}$CO(1--0) emission from NGC 6240
along the double nuclei (position angle is 20$^\circ$).
The abscissa is the angular distance from the $^{12}$CO emission peak
($\alpha=$\timeform{16h50m27s.85}, $\delta=$\timeform{+2D28'57''.9}; B1950).
The ordinate is the velocity offset relative to the systemic velocity of 7339 km s$^{-1}$.
The contour levels are $-2$, 2, 3, 4, 5, 6$\sigma$,
where $1\sigma = $20 mJy beam$^{-1}$ (negative contours are dashed).
The black rectangle represents a single resolution element in both dimensions.
}
\label{copvmap}
\end{figure}

\begin{figure}
\begin{center}
\FigureFile(80mm,220mm){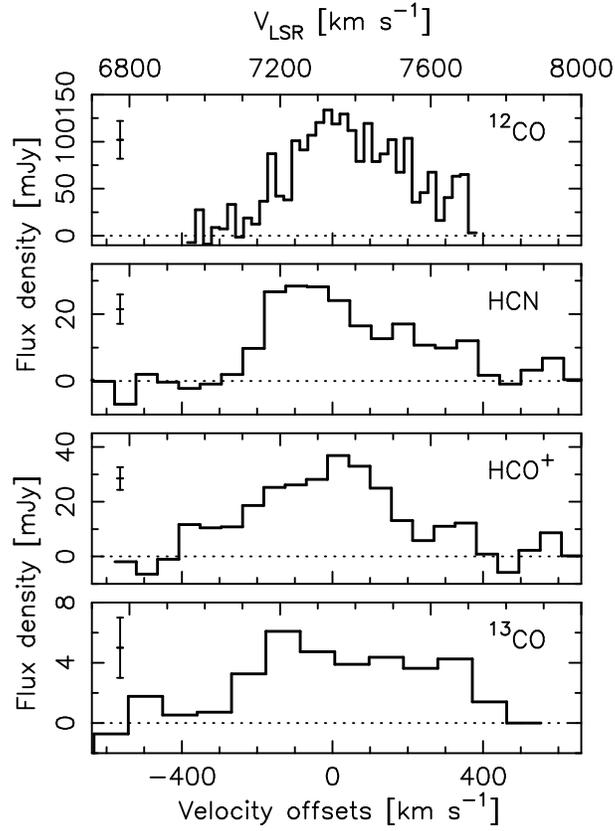}
\end{center}
\caption{
Line profiles of $^{12}$CO(1--0) (top panel), HCN(1--0) (second from top), 
HCO$^+$(1--0) (third from top), and $^{13}$CO(1--0) (bottom) 
at the position of the $^{12}$CO(1--0) emission peak of NGC 6240.
In each panel, the abscissa is labeled with both
the velocity offsets from the systemic velocity of 7339 km s$^{-1}$ 
(lower side) and the LSR velocity (upper side).
}
\label{lineprofs}
\end{figure}

Figure \ref{intmaps}a shows integrated intensity map of 
the $^{12}$CO(1--0) emission.
(No primary beam correction has been applied to the maps 
presented in this paper.)
The distribution of the $^{12}$CO emission is more extended than the 
size of the synthesized beam, suggesting that it is mildly resolved.
The peak of the $^{12}$CO emission is located 
between the double nuclei defined by the 15 GHz continuum peaks
\citep{Carral1990}, not on either of them.
Figure \ref{cochmap} shows the $^{12}$CO velocity channel maps,
and the $^{12}$CO position--velocity map sliced along the double nuclei
(the position angle is \timeform{20D} from north to east)
is presented in figure \ref{copvmap}.
The velocity profile of the $^{12}$CO line at the emission peak is 
presented in figure \ref{lineprofs},
together with those of the HCN, HCO$^+$, and $^{13}$CO emissions.
These plots reveal that the velocity width of the $^{12}$CO line
is extremely large. 
The full width at half maximum is 420 km s$^{-1}$,
and the full width at zero intensity (FWZI) is $>$ 600 km s$^{-1}$
[the recent interferometric observations \citep{Bryant1999,Tacconi1999}
show that the FWZI of $^{12}$CO emission is about 900 km s$^{-1}$].
The observed total integrated intensity of $^{12}$CO is 168.3 Jy km s$^{-1}$,
which is 71\% of the single dish flux 
obtained by the IRAM 30 m telescope (\cite{Solomon1992}).\footnote{
Because of the limitation of bandwidth with the old-FX correlator,
continuum subtraction before imaging could not be performed on 
the $^{12}$CO data.
In order to obtain the continuum-free $^{12}$CO intensity, 
the continuum emission contribution is removed numerically.
The way we employed is as follows.
From the velocity integrated map within 767 km s$^{-1}$ 
(figure \ref{intmaps}a), we primarily obtained the total emission flux 
($^{12}$CO + continuum) of 176.9 Jy km s$^{-1}$.
A 112 GHz continuum flux of 11.2 mJy was estimated from the spectrum 
assuming a single power-law ($S_\nu\propto\nu^{-0.81}$) 
between cm-wave and mm-wave (see also figure \ref{contsed}).
Thus the total continuum emission contribution in figure 
\ref{intmaps}a is estimated to be 8.59 Jy km s$^{-1}$ 
(= 11.2 mJy $\times$ 767 km s$^{-1}$).
We then subtracted this from the primarily flux,
and obtained the continuum-free and velocity- and spatially-integrated
$^{12}$CO line flux of 168.3 Jy km s$^{-1}$.
}
A slight velocity gradient along the double nuclei is seen in 
the position--velocity map (see figure \ref{copvmap}).
The trend of the velocity gradient is the same as 
that in the $^{12}$CO(2--1) map by \citet{Tacconi1999};
the redshifted velocity components are in the north-east, 
and the blueshifted velocity ones are in the south-west.
These results are consistent with the recent $^{12}$CO(1--0) 
observations by Bryant and Scoville (1999).

Figure \ref{intmaps}b shows the HCN(1--0) integrated intensity map.
The continuum emission, which was detected by averaging line-free channels
with an effective bandwidth of 376 MHz, was subtracted from the visibility
data before creating the HCN synthesized image.
Since the HCN emission is unresolved with our $2''$ beam,
it must be more spatially compact than the $^{12}$CO distribution.
The emission peak is located between the double nuclei, 
within \timeform{0''.2} of the peak position of the $^{12}$CO(1--0) emission.
It should be noted that the HCN emission peak (and also $^{12}$CO)
is not coincident with the current star-forming regions traced by 
near-infrared (NIR) emission lines;
the star-forming regions are associated with both of the double nuclei
(e.g. \cite{vanderWerf1993, AlonsoHerrero2002}).
We will focus on this issue in section 5.
The detected total flux of HCN emission is 14.1 Jy km s$^{-1}$,
or $\sim70$\% of single-dish flux obtained by 
Solomon, Downes, and Radford (1992).
The observed flux also agrees very well with the interferometric data
by \citet{Tacconi1999}.
Figure \ref{intmaps}c shows the integrated intensity map of 
HCO$^+$(1--0) emission.
The procedure of continuum subtraction is the same as that of
the HCN data, because both lines are observed simultaneously
in the same passband.
This is the first detection of the HCO$^+$ line from NGC 6240.
Like the $^{12}$CO and HCN emissions, 
the HCO$^+$ emission is peaked between the double nuclei.
The spatial distribution of the HCO$^+$ emission is also compact
and resembles that of the HCN emission.
No systematic velocity gradient is evident in either 
the HCN or the HCO$^+$ data-cubes.
This is probably due to the fact that we have not resolved their spatial 
distributions.
Both the HCN and HCO$^+$ emission-line profiles (figure \ref{lineprofs})
agree reasonably well with that of $^{12}$CO emission, taking into account
their noise levels (20 mJy beam$^{-1}$ per channel for $^{12}$CO,
and 4 mJy beam$^{-1}$ for HCN and HCO$^{+}$).

An integrated intensity map of $^{13}$CO(1--0) emission is presented 
in figure \ref{intmaps}d, and emission is detected at the 4.1$\sigma$ level.
This is the first image ever of $^{13}$CO emission from NGC 6240.
The continuum emission was subtracted from the visibility
data before creating the synthesized image.
The flux of the $^{13}$CO emission is much smaller than 
those of $^{12}$CO and HCN.
The obtained total flux is about 60\% of the single dish flux
obtained by \citet{Casoli1992}.
The synthesized beam size of the $^{13}$CO map 
(\timeform{4''.6}$\times$\timeform{3''.5}) is larger than those of $^{12}$CO
and HCN, and the emission dose not seem to be resolved.
The peak of $^{13}$CO also coincides with those of $^{12}$CO, HCN, and HCO$^+$.

\subsection{Line Intensity Ratios}

In this subsection, we briefly comment on the emission line intensity ratios,
in brightness temperature scale,\footnote{
The relation between line flux ($S$) and 
brightness temperature ($T_{\rm b}$) is 
$T_{\rm b} = S\cdot\lambda^2/2k\Omega_{\rm B}$,
where $\lambda$ is the observed wavelength,
$k$ is the Boltzmann constant, and $\Omega_{\rm B}$ is 
the solid angle of the region in which the line ratio is calculated.
}
for the molecular gas concentration between the double nuclei.
Taking into account the relatively low spatial resolution of the $^{13}$CO map,
the line ratio is calculated using the integrated fluxes within the
region of \timeform{4''.6}$\times$\timeform{3''.5} (2.2 kpc $\times$ 1.7 kpc),
which corresponds to the synthesized beam size of the $^{13}$CO map.
This area both contains the region between the double nuclei and 
the double nuclei, themselves.
However, the emissions are almost concentrated between the double nuclei,
and the peaks of the observed emission lines coincide with each other.
Thus, the line ratios within the \timeform{4''.6}$\times$\timeform{3''.5} 
region are representative of those between the double nuclei.
The error of each ratio is calculated considering 
the absolute flux calibration uncertainty for each observation (10\%) and
the noise level of the emission line maps, except for HCO$^+$/HCN.
For the HCO$^+$/HCN ratio, only the noise level of each emission line map 
is considered, because the HCN and HCO$^+$ line is observed
simultaneously within the same passband of the correlator.

We obtained a high HCN/$^{12}$CO ratio of $0.25\pm0.04$ 
at the region between the double nuclei.
This contrasts with the total integrated intensity ratio of 
HCN/$^{12}$CO = $0.14\pm0.02$, which agrees with the single-dish values
reported by Solomon, Downes, and Radford (1992).
The HCN/$^{12}$CO ratio of 0.25 is higher than the mean ($\approx 0.1$) for 
LIRGs and ULIRGs, and one of the highest value among them \citep{Gao2004b}.
Gao and Solomon (2004a) show that a large HCN/$^{12}$CO ($\geq0.1$) is
likely to come from regions where most of ($\gtrsim$ 50\%) molecular gas is 
high density [$n{\rm (H_2)}>10^4$ cm$^{-3}$] based on LVG calculations.

We obtain a HCO$^+$/HCN ratio of $1.5\pm0.1$.
The typical HCO$^+$/HCN ratio for nearby starburst galaxies 
and Seyfert galaxies with nuclear starbursts is 0.3 -- 2
(\cite{Nguyenqrieu1992, Kohno2001}).
Therefore, the value in NGC 6240 is within the range for that of
starburst galaxies, but larger than the mean value ($\lesssim1$).
The high HCO$^+$/HCN ratio would be due to the existence of shocked dense gas,
of which the envelope is exposed to the intense ionization flux 
of the supernovae (\cite{Nguyenqrieu1992}, and references therein).
The HCO$^+$/HCN ratio of this galaxy will be discussed in a forthcoming paper
together with those in the other LIRGs and mergers.

The $^{12}$CO/$^{13}$CO ratio is $21\pm6.2$
at the region between the double nuclei (2.2 kpc $\times$ 1.7 kpc).
\citet{Aalto1995} found that the mean $^{12}$CO/$^{13}$CO ratio is
$\sim12$ in their survey of infrared-bright galaxies.
The ratio in NGC 6240 is thus significantly larger 
than the mean for starburst galaxies.
Some galaxies (e.g., NGC 1614, IC 694) show a high ratio of 
$^{12}$CO/$^{13}$CO$>20$, 
and \citet{Aalto1995} suggested that such a high ratio could be reproduced 
by warm turbulent molecular gas ($T>100$ K).
They also maintain that a high $^{12}$CO/$^{13}$CO ratio also implies that 
the optical depth of $^{12}$CO(1--0) must be moderate [$\tau$($^{12}$CO)$\sim1$]
and that of $^{13}$CO(1--0) is very small [$\tau$($^{13}$CO)$\ll1$].

\subsection{Properties of the Continuum Emission}

\begin{figure}
\begin{center}
\FigureFile(160mm,220mm){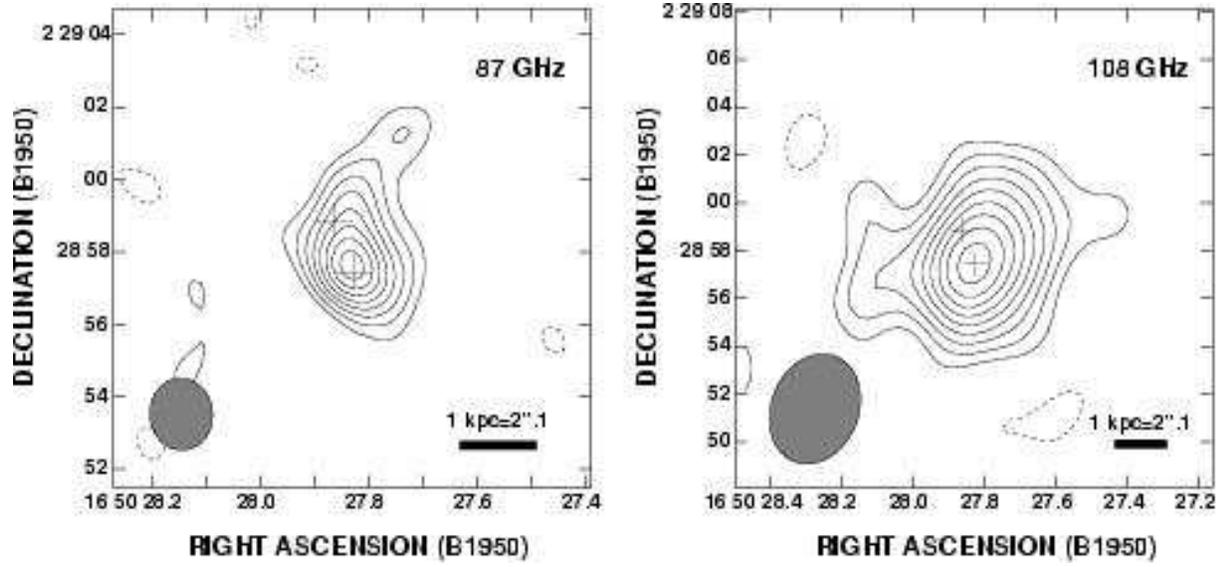}
\end{center}
\caption{
Continuum maps at 87 GHz (left panel) and 108 GHz (right panel)
toward the center of NGC 6240.
The contour levels are $-2$, 2, 3, 4, ..., 9$\sigma$,
where $1\sigma =$ 1.0 mJy for the 87 GHz map,
and $-2$, 2, 3, 4, ..., 10$\sigma$, where $1\sigma =$ 0.7 mJy
for the 108 GHz map (negative contours are dashed).
The synthesized beam is shown at the bottom-left corner of the map.
The crosses show the peak positions of the 15 GHz continuum (\cite{Carral1990}).
No primary beam correction has been applied.
}
\label{contmap}
\end{figure}

Figure \ref{contmap} shows both the 87 GHz (3.5 mm) and 108 GHz (2.8 mm)
continuum images obtained using the line-free channels of 
the visibility data in HCN$/$HCO$^+$ and $^{13}$CO observations, respectively.
Unlike the molecular lines, the continuum emission peak is almost located
at the southern nucleus.
Since the HCN and HCO$^+$ maps and 87 GHz continuum map were created from
identical uv-data, the difference between the peak positions is real.
In the 87 GHz continuum map, the emission extends more than 
the size of the synthesized beam (\timeform{3''.6}$\times$\timeform{2''.2},
which corresponds to 1.7 kpc$\times$1.0 kpc).

\begin{figure}
\begin{center}
\FigureFile(80mm,220mm){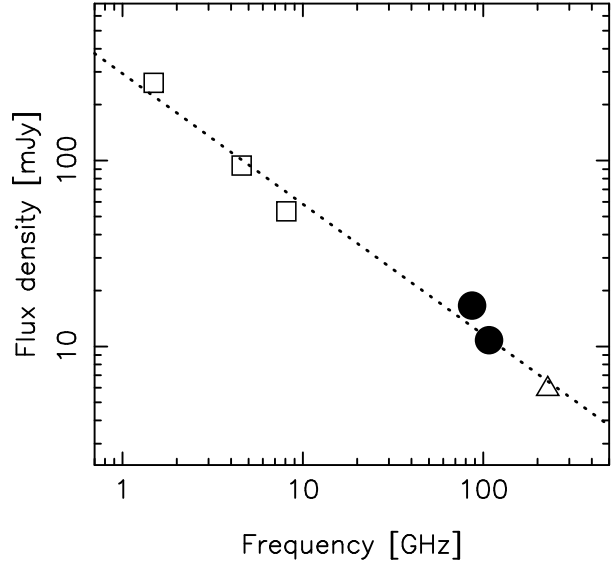}
\end{center}
\caption{
The cm- to mm-wave continuum spectrum of
NGC 6240 nuclear region within \timeform{10''.8}$\times$\timeform{13''.6}.
Data are from \citet{Colbert1994} (open squares), \citet{Tacconi1999}
(open triangle), and this work (filled circles).
The errors for each data point are smaller than 
the sizes of the symbols in the plot.
The dotted line represents the best fit power-law with an spectral 
index of $-0.81$.
}
\label{contsed}
\end{figure}

Figure \ref{contsed} shows the spectrum
of NGC 6240 in the region surrounding the double nuclei
(within \timeform{10''.8}$\times$\timeform{13''.6}, 
which corresponds to 5.1 kpc$\times$6.5 kpc).
The total continuum flux densities at 87 GHz and 108 GHz are
16.6 mJy and 10.8 mJy, respectively. 
These values can be reproduced well by assuming
a single power-law spectrum ($S_\nu\propto\nu^\alpha$)
interpolated between the cm-wave continuum emission \citep{Colbert1994}
and 1.3 mm (228 GHz) emission \citep{Tacconi1999}.
The best-fit spectral index is $\alpha=-0.81$, 
which is within the range of the spectral indices in both synchrotron radiation
from AGN and that from supernova remnants (SNRs)
(e.g., \cite{Maslowski1984,Condon1992}).
Therefore, most of our observed mm-wave continuum emission can be
explained by synchrotron radiation from an AGN, SNRs, or both, 
come mostly from the bright southern nuclei,
but there are surely some contributions from the northern nucleus.
This conclusion supports the discussion by \citet{Tacconi1999}.

Thermal free--free emission from star-forming regions are also expected 
to contribute some degree to the mm-wave continuum emission.
Based on \citet{Condon1992} and \citet{Kennicutt1998}, the relation of 
the thermal free--free continuum flux ($S_{\rm free-free}$)
and far-infrared luminosity ($L_{\rm FIR}$) is expressed 
as follows:
\begin{equation}
\frac{S_{\rm free-free}}{\rm mJy}
=2.28\times10^{-7}\times\Bigl(\frac{D}{\rm Mpc}\Bigr)^{-2}
\times\Bigl(\frac{T_{\rm e}}{10^4{\rm K}}\Bigr)^{0.59}
\times\Bigl(\frac{\nu}{\rm GHz}\Bigr)^{-0.1}
\times\frac{L_{\rm FIR}}{\LO},
\end{equation}
where $T_{\rm e}$ is the electron temperature.
Even if we assume all of the FIR luminosity from NGC 6240,
$L_{\rm FIR}=4.9\times10^{11}$ \LO\ \citep{Sanders1991},
comes from the nuclear region, 
then the estimated free--free emission flux from 
H {\sc ii} regions at $T_{\rm e}=10^4$ K is at most $\sim 7$ mJy at 100 GHz.
This flux value is smaller than the observed flux at 87 GHz and 108 GHz.
Moreover, as can be seen in figure \ref{contsed},
the spectral index between 100 GHz and 228 GHz is 
significantly smaller than $-0.1$, which is a typical index for 
optically thin radio free--free emission. 
Therefore, the free-free emission contribution on 
the 100 GHz continuum flux must be less than 7 mJy.


\section{Dense and Warm Molecular Gas between the Double Nuclei of NGC 6240}

Intensity ratios between molecular emission lines 
are powerful tools to probe the physical properties of molecular gas.
In this section we compare the observed HCN/$^{13}$CO intensity ratio
with radiative transfer model calculations,
to try to constrain the physical properties of 
the molecular gas between the double nuclei of NGC 6240.

We have compared the observed HCN/$^{13}$CO ratio with the outcome of
radiative transfer model calculations performed by \citet{Matsushita1998}
based on the large velocity gradient 
(LVG) approximation \citep{Goldreigh1974, Scoville1974}.
The intense emission of the HCN line and high HCN/$^{12}$CO ratio (0.25)
imply that most of the molecular gas is dense between 
the double nuclei \citep{Solomon1992,Gao2004a}.
HCN emission traces molecular gas at high density
[$n({\rm H_2}) > 10^{4}$ cm$^{-3}$],
and $^{13}$CO emission is more excited at higher density gas than $^{12}$CO
due to its low optical depth.
Thus, the HCN/$^{13}$CO ratio reflects the physical properties
of dense molecular gas better than HCN/$^{12}$CO ratio \citep{Matsushita1998}.

\begin{figure}
\begin{center}
\FigureFile(120mm,220mm){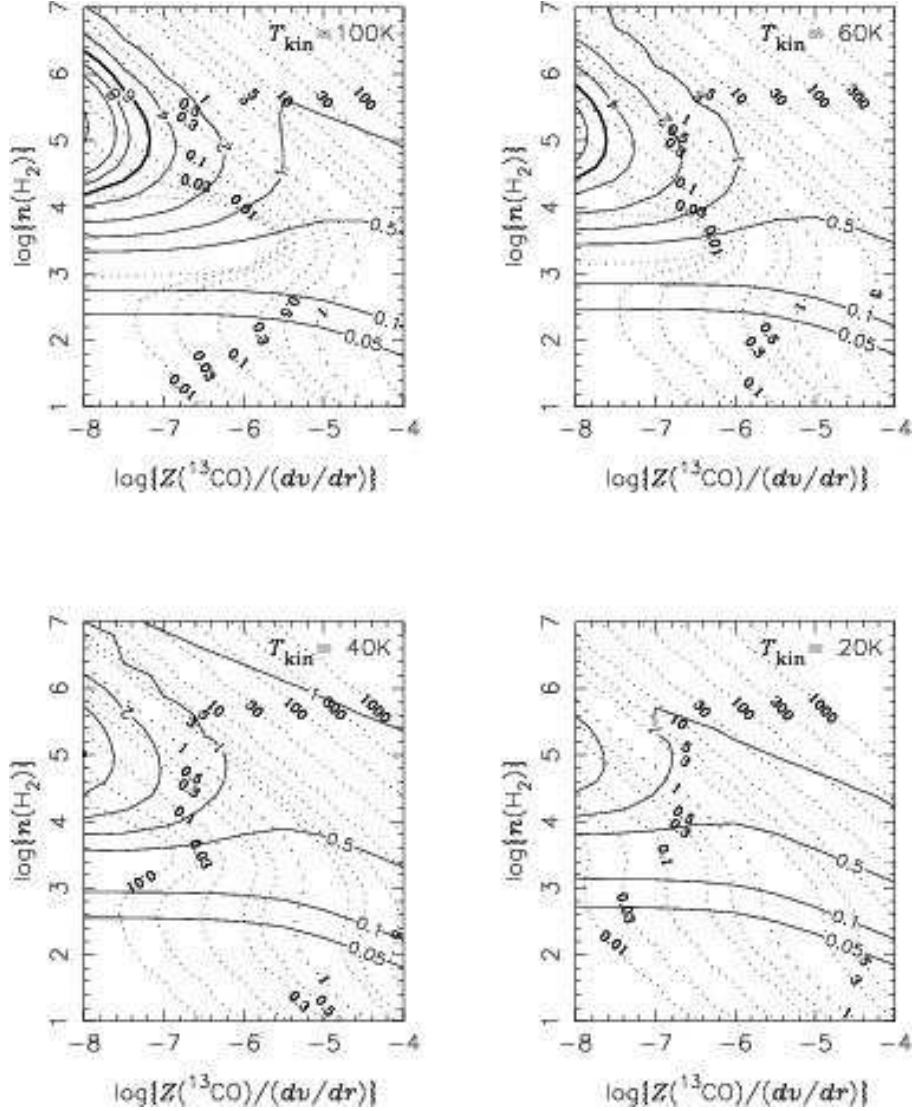}
\end{center}
\caption{
Simulated HCN/$^{13}$CO intensity ratio (solid contour) and
opacity of $^{13}$CO emission [$\tau_{1-0}$($^{13}$CO); dotted contour]
obtained from LVG calculations by \citet{Matsushita1998}.
A fixed relative abundance of [$^{13}$CO]/[HCN]=50 is assumed.
Models with kinetic temperatures ($T_{\rm kin}$) of 20, 40, 60,
and 100 K are presented 
($T_{\rm kin}$ is indicated in the top right corner of each panel).
The ordinate is $Z(^{13}{\rm CO})$/($dv/dr$),
where $Z(^{13}{\rm CO})$ is the relative abundance of $^{13}$CO to the H$_2$
and $dv/dr$ is the velocity gradient in km s$^{-1}$ pc$^{-1}$.
The abscissa is the number density of molecular hydrogen in cm$^{-3}$.
The contour levels in solid lines (HCN/$^{13}$CO) are 
0.05, 0.1, 0.5, 1, 2, 4, 6, 8, 10, and 15. 
The thick solid curve indicates a HCN/$^{13}$CO ratio of 6.
The contour levels in dotted lines [$\tau_{1-0}$($^{13}$CO)] are 
0.01, 0.03, 0.1, 0.3, 0.5, 1.0, 3.0, 5.0, 10.0, 30.0, 100.0, 300.0, and 1000.0.
The opacity of unity are shown as thick dotted curves.
}
\label{lvgreslt}
\end{figure}

Figure \ref{lvgreslt} shows the contour maps of the intensity ratio,
HCN/$^{13}$CO, as a function of $Z(^{13}{\rm CO})$/($dv/dr$) and $n({\rm H_2})$,
where $Z(^{13}{\rm CO})$ is the fractional abundance of $^{13}$CO to H$_2$
and $dv/dr$ is the velocity gradient in km s$^{-1}$ pc$^{-1}$.
The kinetic temperature ($T_{\rm kin}$) in each map is
shown in the top-right corner.
In the LVG calculation, the single-component model was employed,
and a constant relative abundance of [$^{13}$CO]/[HCN] = 50 
was adopted \citep{Solomon1979,Irvine1987}.
The full description of the LVG calculation is presented in 
\citet{Matsushita1998}.

As can be seen in figure \ref{lvgreslt},
the observed high ratio, HCN/$^{13}$CO = $5.9\pm1.7$,
can be reproduced in dense and warm molecular clouds.
In the upper two panels showing kinetic temperatures higher than 50 K,
the ratio can only be reproduced by dense gas conditions,
$n({\rm H_2}) > 10^4$ cm$^{-3}$.
The optical depth of $^{13}$CO emission under the dense and
warm conditions is very small (0.01 -- 0.1)
in the panels of $T_{\rm kin}=60$ K and 100 K, 
as suggested by \citet{Aalto1995}.

To explain the high HCN/$^{13}$CO ratio under 
a lower gas kinetic temperature ($T_{\rm kin} \leq 40$ K),
extremely small values of $Z(^{13}{\rm CO})$/($dv/dr$)
[$<10^{-8}$ (km s$^{-1}$ pc$^{-1}$)$^{-1}$] would be required.
This would happen if 
the velocity gradient is extremely large
($dv/dr > 10^2$ km s$^{-1}$ pc$^{-1}$), or
the $^{13}$CO abundance is very low [$Z(^{13}{\rm CO})<10^{-8}$].

If a $^{13}$CO abundance of $Z(^{13}{\rm CO}) =1\times10^{-6}$ 
(\cite{Solomon1979}) is adopted, a velocity gradient of 
$dv/dr > 10^2$ km s$^{-1}$ pc$^{-1}$ is needed to meet
the extremely small $Z(^{13}{\rm CO})$/($dv/dr$) value.
As we show now, this velocity gradient is, however, too large and unrealistic.
We made a simple estimate of the velocity gradient from
the $^{12}$CO emission data.
\citet{Tacconi1999} reported that, at the peak of $^{12}$CO(2--1) emission,
the line width at zero intensity (FWZI) is about 1000 km s$^{-1}$
in a \timeform{0''.7} $\times$ \timeform{0''.5} (= 330pc $\times$ 240pc) beam.
Assuming that the $^{12}$CO emission originated from a single molecular clump
having the same size as the observed beam, the velocity gradient is
nominally estimated to be $dv/dr \sim 3$ km s$^{-1}$pc$^{-1}$.
Even if the size of molecular clump were 50 pc 
(typical for Galactic giant molecular clouds),
the estimated velocity gradient would be $dv/dr \sim 18$ km s$^{-1}$ pc$^{-1}$,
which is still quite smaller than the value above.
Therefore, the extremely large velocity gradient as
$dv/dr > 10^2$ km s$^{-1}$ pc$^{-1}$ is not realistic.

We now discuss the possibility that the $^{13}$CO abundance is quite low.
In the calculations for figure \ref{lvgreslt},
a constant relative abundance of [$^{13}$CO]/[HCN]=50 was used.
The HCN/$^{13}$CO ratio hardly depends on the HCN abundance
[$Z({\rm HCN})$/($dv/dr$)] because of the large optical depth of 
HCN (see figure 6 of \cite{Matsushita1998}).
On the other hand, a low $^{13}$CO abundance can reproduce high 
HCN/$^{13}$CO ratios.
The ``standard abundance'' $Z(^{13}{\rm CO}) =1\times10^{-6}$ is obtained
based on observations of molecular clouds in the inner Galaxy
(\cite{Solomon1979}).
Although the $^{13}$CO abundance at the centers of other galaxies
are still not very well constrained, 
there appears to be little difference between the Galaxy and others
(e.g. \cite{Henkel1993},\yearcite{Henkel1994},\yearcite{Henkel1998}).
The galaxy--galaxy merging may cause a low $^{13}$CO abundance
in the galactic center.
The molecular gas in the disk of merging progenitors are predicted 
to converge into the centers of each of the original galaxies
\citep{Barnes1996, Mihos1996}.
In the Galaxy, the $^{12}$C/$^{13}$C abundance ratio becomes 
larger toward the outer disk than in the Galactic center,
but the difference is only a factor of $2\sim5$
(\cite{Wilson1994}, and references therein; \cite{Savage2002}).
The $^{12}$CO/$^{13}$CO abundance ratio will not largely differ from 
that of the $^{12}$C/$^{13}$C (e.g. \cite{Langer1984}).
Thus, the low $^{13}$CO abundance might somewhat contribute to 
the high HCN/$^{13}$CO ratio, but the extremely low $^{13}$CO abundance 
[$Z(^{13}{\rm CO})<10^{-8}$] cannot be realized.
Selective dissociation of the $^{13}$CO molecule might play a role,
but the abundance of $^{13}$CO is supposed to be affected little
\citep{vandishoeck1988}.

From the above considerations, it seems to be quite reasonable to conclude 
that the molecular gas is dense [$n({\rm H}_2) = 10^{4-6}$ cm$^{-3}$]
and warm ($T_{\rm kin}>50$ K) between the double nuclei of NGC 6240.
Our results are also consistent with the fact that 
the analysis of the spectral energy distributions
from the infrared to the sub-mm suggests 
a significant amount of warm ($T_{\mathrm{dust}}\gtrsim50$ K) dust
in NGC 6240 (\cite{Klaas1997, Lisenfeld2000}).
To confirm the high-temperature condition of the molecular gas positively, 
observations of higher excitation $^{12}$CO and $^{13}$CO emission
should be performed.
It is expected that the $^{12}$CO/$^{13}$CO ratio in $J=$2--1 and 3--2 show 
values smaller than that of $J$=1--0 (cf. \cite{Aalto1995,Huttemeister2001}).

Here, we make a brief comparison of physical properties of molecular gas 
with those of other starburst galaxies.
The observational results of higher excitation $^{12}$CO emission lines
($J=$7--6/6--5) combined with LVG calculations suggest that, 
for the central region ($r\lesssim 300$ pc),
the physical properties of molecular gas is 
$n({\mathrm{H}_2})\gtrsim10^4$ cm$^{-3}$ and $T_{\rm kin}\gtrsim100$ K 
in NGC 253 \citep{Bradford2003} and M 82 \citep{Mao2000,Ward2003}.
\citet{Aalto1997} obtained interferometric maps of 
$^{12}$CO, $^{13}$CO, and HCN (1--0) toward the LIRG merger 
Arp 299 (IC 694+NGC 3690) 
in \timeform{2''.3} -- \timeform{5''.4} (460 -- 1090 pc) resolutions.
At the nucleus of IC 694, the ratio $^{12}$CO/$^{13}$CO is 60, 
the ratio HCN/$^{12}$CO is 0.11, and the ratio HCN/$^{13}$CO is 6.7.
These ratios are similar to those of NGC 6240.
Single-dish $^{13}$CO(2--1) observations toward IC 694 have also been
performed by \citet{Aalto1995}; 
they concluded that the molecular gas consists of a dense 
[$n({\rm H}_2)=10^{4-5}$ cm$^{-3}$] and warm ($T_{\rm kin}>50$ K) medium
based on a high $^{13}$CO(2--1)/(1--0) ratio ($\gtrsim2$).
These previous results imply that the estimated properties of 
the molecular gas in NGC 6240 are common for starburst and merging galaxies.


\section{Discussion}

\subsection{Dense Molecular Gas and Star Forming Regions}

As we have shown in previous sections, NGC 6240 has a large amount of
dense and warm molecular gas in the central region.
The dense molecular gas distribution, however, 
does not coincide with that of the massive star-forming regions.

\begin{figure}
\begin{center}
\FigureFile(160mm,220mm){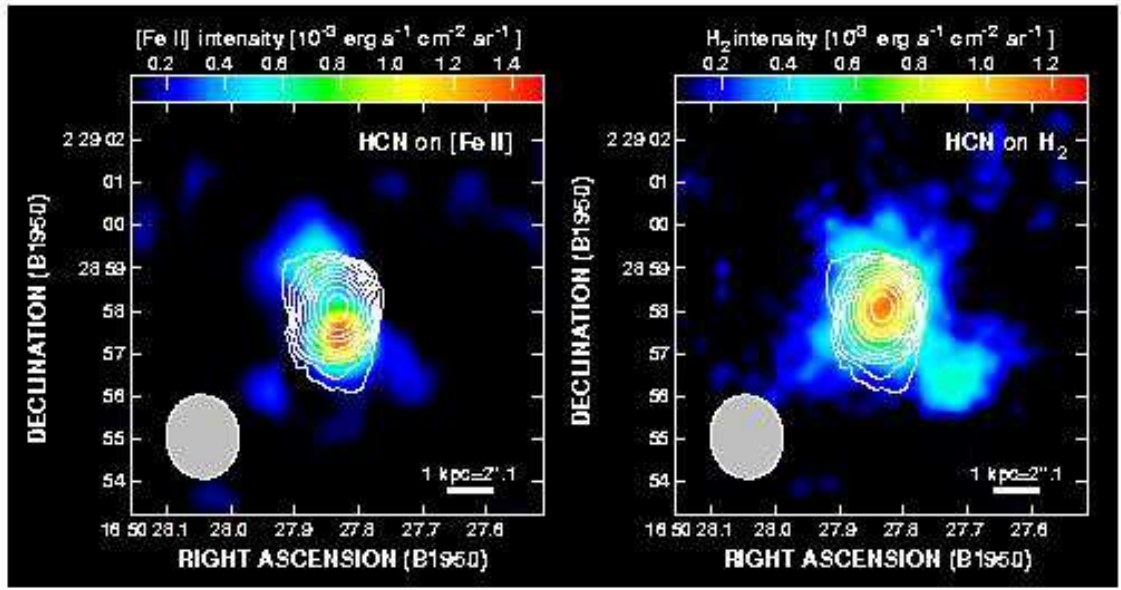}
\end{center}
\caption{
HCN integrated intensity map (contours) superimposed on the near-infrared
[Fe {\sc ii}] at $1.644$ $\mu$m (left panel) and 
H$_2$($v=$1--0 $S(1)$) at $2.121$ $\mu$m (right panel) emission line maps
from \authorcite{vanderWerf1993}(\yearcite{vanderWerf1993}, 2005 in preparation).
The contour levels for HCN are the same as figure 1b.
The synthesized beam of the HCN is shown at the bottom-left corner.
}
\label{hcnonfeii}
\end{figure}

Figure \ref{hcnonfeii} shows our HCN map superimposed on the distribution 
of NIR [Fe {\sc ii}] (left panel) and H$_2$ (right panel) emissions.
The figure clearly shows that the HCN peak agrees well
with the H$_2$ peak rather than the [Fe {\sc ii}] peak.
The [Fe {\sc ii}] emission is supposed to be generated in
fast shocks from supernova remnants \citep{vanderWerf1993,Sugai1997},
and is thus supposed to follow the current massive star forming regions.
On the other hand, the H$_2$ emission arise from hot molecular hydrogen
excited by large-scale shock rather than UV fluorescence 
or X-ray heating (\cite{Sugai1997,Ohyama2003}).
The dense gas traced by the HCN and HCO$^+$ emissions are 
concentrated between the double nuclei (see also figure \ref{intmaps}), 
whereas the massive star-forming regions traced by NIR [Fe {\sc ii}], 
Pa$\alpha$, and Br$\gamma$ emission lines 
are associated with both of the nuclei
(\cite{vanderWerf1993, AlonsoHerrero2002, Tecza2000}).
Usually the dense molecular gas is located at current massive 
star-forming regions in the central region of starburst galaxies 
and Seyferts \citep{Kohno1999,Shibatsuka2004}.
This is in contrast with what we have observed in NGC 6240.
We provide a possible illustration to explain this discrepancy 
in the next subsection.

\subsection{Merging Evolution of Dense Molecular Gas and Star Formation in NGC 6240}

Here, we propose a possible scenario to explain the evolution of 
the molecular gas and star formation in NGC 6240.
The origin of the unusual discrepancy between the dense gas distribution 
and the star-forming regions is also illustrated within it.
This scenario is based on our observational results together 
with those of others and results from numerical simulations of 
galaxy--galaxy merging.
A brief summary of the scenario is as follows:
1) during a galaxy--galaxy interaction,
dense molecular gas is formed and intense star formation occurs
in the circumnuclear regions of each of the merger progenitor galaxies,
2) an almost head-on collision or very close encounter of the nuclei cause
the merging of the two dense gas concentrations,
3) the star-forming nuclei can still remain separate while
the single dense gas concentration is left behind between them
(this is the present view of NGC 6240),
4) a new and intense star formation will begin at some period 
in the dense molecular gas concentration.

We describe this scenario in more detail following the order of events.

\subsubsection{Pre-collision phase: gas accumulation and circumnuclear
star formation on each merger progenitor}

At the early merging stages, the molecular gas pre-existing
in the merger progenitor galaxies accumulate
toward their respective galactic centers approaching each other.
The gas accumulation probably results in the formation of dense molecular gas
and nuclear starburst detected by the NIR observations (e.g. \cite{vanderWerf1993}).
N-body simulations of galaxy--galaxy merging show that
gas inflow, driven by gravitational torques, will cause condensations of 
gas clouds around the nuclei of both original galaxies
\citep{Barnes1996, Mihos1996, Barnes2002}.
Numerical simulations also predict that the gas accumulation in each of 
the nuclei probably lead to considerable amount of dense molecular gas formation,
which is able to cause star-formation activities within it.
For example, in the merging galaxy Arp 299, 
most of the molecular gas is concentrated 
toward the nuclei \citep{Aalto1997}, and the nuclei harbor
intense star formation activity \citep{AlonsoHerrero2000}.

\subsubsection{Head-on collision and formation of single dense gas concentration}

In the course of the merging process, 
the progenitor nuclei accompanied by dense gas concentration 
may suffer a nearly head-on collision or very close encounter.
In the case that the impact parameter was smaller than the gas concentrations
($<1$ kpc), although the nuclei and associated stellar systems 
can remain gravitationally bounded, 
the dense molecular gas would be stripped off from the nuclei
and form a single molecular gas condensation
(e.g. \cite{Braine2004,Struck1997}).
These events ultimately lead the nuclei to be deficient in dense molecular gas,
while almost terminating the nuclear star formation.

We apply this to NGC 6240, and try to make a rough estimate of 
the time-scale of the collision/encounter using the relative velocity 
and distance between the double nuclei.
The projected relative velocity is
measured to be $\sim 50$ km s$^{-1}$ from the stellar absorption and
$\sim 150$ km s$^{-1}$ from the Br$\gamma$ emission \citep{Tecza2000};
thus, the actual relative velocity is assumed to be on the order of 100 km s$^{-1}$.
From the 15 GHz continuum image \citep{Carral1990},
the projected distance of the double nuclei is 710 pc.
If our viewing angle is \timeform{45D},
then the actual separation of the nuclei is about 1 kpc.
If we adopt 100 km s$^{-1}$ for the relative velocity and 1 kpc for the distance,
it takes about $10^7$ yr for the double nuclei to move 
from the pericenter to the present positions.
\citet{Tecza2000} estimated that the actual orbital velocity and separation 
between the double nuclei are 155 km s$^{-1}$ and 1.4 kpc, respectively.
They are very close to our assumed values.
This time-scale is consistent with the age of the nuclear star formation.
The latest star-formation activity is estimated to be suspended for
1--1.5 $\times 10^7$ yr ago from the results of 
the stellar population synthesis 
based on the optical and NIR spectroscopic observations
(\cite{Schmitt1996, Tecza2000}).

\subsubsection{Post-collision phase: present view of NGC 6240}

After collision/encounter,
the double nuclei leave each other from the pericenter, and
the dense molecular gas concentration remains between the double nuclei.
Some recently formed massive stars exist around the nuclei,
but the major part of nuclear star-formation activities have already faded out 
because of the deficiency of dense molecular gas.
This is the present view of NGC 6240.

The molecular gas concentration shows a velocity gradient within it;
the north side is redshifted with respect to the south 
(see figure \ref{copvmap}).
This is consistent with the orbital motions of the double nuclei
\citep{Tecza2000}.
\citet{Tacconi1999} concluded that the velocity gradient originates from
the rotational motion of the gas concentration.
In a perfect head-on collision, the merged gas is expected to 
have no systematic velocity structure.
But, in the case of an offset head-on collision or a very close encounter,
the orbital angular momentum of dense gas in the pre-collision phase 
will remain as the rotational motion of the merged gas concentration.
In our scenario, the observed velocity gradient can be explained 
as such rotational motion.

The molecular gas between the double nuclei is not only dense, but also warm.
There is no sign of intense massive star formation or prominent AGN activity
between the double nuclei,
and thus ultraviolet radiation from them should not be 
a principal heating source.
One probable heating source is dynamical energy sources, 
that is to say, turbulence and shocks.
The large velocity width of the molecular emission lines and 
complicated distributions of molecular gas imply the existence of 
large turbulence within the dense gas concentrations \citep{Tacconi1999}.
Extraordinary bright shock excited NIR H$_2$ emission has been detected
in this object (\cite{Sugai1997,Ohyama2000,Gerssen2004}),
and the H$_2$ peak agrees well with that of the dense gas concentration 
(see figure \ref{hcnonfeii}).
These facts imply that the large turbulence and the intense shocks
causes a high molecular gas temperature.

\subsubsection{New starburst in dense gas concentration}

A brand-new starburst will begin in the dense gas concentration
in the near future.
\citet{Tacconi1999} estimated the dissipation time-scale for random
velocities within the gas concentration to be $\leq 10^7$ yr,
and suggest that NGC 6240 may 
experience the next major starburst within the gas concentration.
Some external effects on the dense gas might also proceed star-formation.
The ``starburst driven starburst'' proposed by \citet{Taniguchi1998}
is one of the possible mechanisms.
Dense gas clumps can collapse under high external pressure caused 
by the starburst driven super-wind from the double nuclei \citep{Ohyama2003},
and massive star-formation will be induced in them.

Star formation within the dense gas concentration
may have already commenced, but it is very young ($\ll 10^7$ yr) and 
suffering from large interstellar absorption because of huge amounts of
interstellar medium.
There exists a hint that the starburst activity should not be older 
than several $10^6$ yr, if the starburst has already begun.
There is no evident sign of strong supernova explosions,
because of the weakness of cm-wave radio continuum emission 
and NIR [Fe \emissiontype{II}] emission between the double nuclei
\citep{Colbert1994,vanderWerf1993}.

Mid-infrared (MIR) spectroscopic observations with 
the Infrared Space Observatory (ISO)
revealed that the star-forming regions in NGC 6240 
suffer from large interstellar extinction.
The extinction toward the star-forming regions, estimated from
the emission line flux ratio between MIR [Ne {\sc ii}] and 
NIR Br$\gamma$, is $A_V=$15 -- 20 mag 
(corresponding to $A_K\sim$ 1.5 -- 2 mag)
in the simplified screen assumption (\cite{Lutz2003}).
This estimation is larger (10 -- 15 mag in $A_V$) 
than the extinction derived from optical and NIR spectroscopic observations
(\cite{Veilleux1995, Tecza2000}).
Therefore, the spatial distributions seen even at the NIR wavelength
([Fe {\sc ii}], Pa$\alpha$, and Br$\gamma$ emission lines) may not totally
trace star-forming regions deeply obscured by the interstellar medium.
Although the location of an obscured star-forming region has not been 
identified yet because the ISO aperture ($> 10''$) covers 
the whole nuclear region of NGC 6240, a young star-forming region 
may hide in the dense molecular gas concentration between the double nuclei.


\section{Summary}

We have carried out high spatial-resolution observations of molecular lines
toward the infrared luminous merger NGC 6240
using the NMA and the RAINBOW interferometers.
The $^{12}$CO emission peak lies between the double nuclei of this galaxy;
our results confirm the previous interferometric $^{12}$CO(1--0/2--1)
observations.
Our new high-resolution HCN and HCO$^+$ observations reveal that
the emissions are also peaked between the double nuclei,
and are not spatially resolved with our beam size of \timeform{2''}.
The ratio HCN/$^{12}$CO is as high as 0.25 
in the region between the double nuclei.
The $^{13}$CO emission has also been observed, 
and it turns out to be much weaker than others.

The comparison between the high ratio HCN/$^{13}$CO = 5.9 and 
the LVG calculations 
suggests that the molecular gas is substantially dense and warm.
The estimated molecular hydrogen density and kinetic temperature are
$n({\rm H_2})=10^{4-6}$ cm$^{-3}$ and $T_{\rm kin}>50$ K, respectively.
These physical properties are similar to
those seen at the centers of the nearby starburst galaxies
and infrared luminous mergers.

The observed characteristics that most of the dense and warm gas in NGC 6240
is located between the starburst nuclei is very unusual 
among the infrared luminous mergers observed so far.
We propose a scenario to explain the evolution of 
dense molecular gas and starbursts, and to resolve
the discrepancy between their locations in NGC 6240.
In our scenario, a head-on collision or very close encounter of 
the star-forming nuclei caused the merging of dense molecular gas
associated with both nuclei.
At the present moment, the double star-forming nuclei are still apart from
each other, but a dense and warm molecular gas concentration has been left
between them.
In the near future, a new starburst is expected to commence 
within the dense gas concentration.

\bigskip

The authors wish to express their deep gratitude to Dr. Satoki Matsushita 
for kindly providing us with his LVG calculation results.
We are grateful to Prof. Paul van der Werf for readily furnishing us
with his high-quality near-infrared images of NGC 6240.
We are indebted to Dr. Baltasar Vila-Vilar\'{o} for carefully reading 
the manuscript and valuable comments.
We would like to thank Dr. Dennis Downes, the referee, for his helpful comments.
We thank the NRO staff for the operation of the telescopes 
and assisting us in our observations.
K.N. was financially supported by the Japan Society for 
the Promotion of Science (JSPS).
\ K.N. was also partially supported by a Grant-in-Aid for Scientific Research
from the Ministry of Education, Culture, Sports, Science and Technology (MEXT) No. 15037212.
\ K.K. was financially supported by 
JSPS Grant-in-Aid for Scientific Research (B) No.14403001 and 
MEXT Grant-in-Aid for Scientific Research on Priority Areas No.15071202.
The Nobeyama Radio Observatory is a branch of 
the National Astronomical Observatory of Japan, 
the National Institutes of Natural Sciences (NINS).



\clearpage

\begin{table}
\begin{center}
\caption{Observational log.}\label{obslog}
\begin{tabular}{lcccc} \hline\hline
 & $^{12}$CO & HCN & HCO$^+$ & $^{13}$CO \\ \hline
 Array configurations & AB,C,D & Rb\footnotemark[$*$], AB, C & Rb, AB, C & C, D \\
 Observed frequency (GHz) & 112.52 & 86.52 & 87.06 & 107.57 \\
 Year of observation & 1995--1996 & 2000--2002 & 2000--2002 & 2001--2003 \\
 Correlator & old-FX & UWBC & UWBC & UWBC \\
 Frequency resolution (MHz)& 0.3125 & 8 & 8 & 8 \\
 Bandwidth (MHz)& 320 & 1024 & 1024 & 1024 \\
 Synthesized beam($''$) & $2.2\times2.1$ & $2.0\times1.7$ & $2.0\times1.7$ & 
$4.6\times3.5$ \\
 \hline
\multicolumn{5}{@{}l@{}}{\hbox to 0pt{\parbox{85mm}{\footnotesize
\par\noindent
\footnotemark[$*$] RAINBOW.
}\hss}}
\end{tabular}
\end{center}
\end{table}

\begin{table}
 \begin{center}
\caption{Observational results.}\label{obsres}
 \begin{tabular}{lcccc} \hline\hline
  & $^{12}$CO & HCN & HCO$^+$ & $^{13}$CO \\ \hline
  On-source integration time (hour) & 9.7 & 17.0 & 17.0 & 27.1 \\
  Velocity resolution of data cube (km s$^{-1}$) & 21.3 & 55.4 & 55.1 & 178.4\\
  Typical noise level (mJy beam$^{-1}$)& 20 & 4 & 4 & 2\\
  Typical T$_{\rm sys}$ in SSB (K) & 760 & 330 & 330 & 400 \\
   &  &  &  &  \\
  Half intensity size ($''$) & $3.0\times2.4$ & $2.0\times1.7$ & $2.3\times2.1$ & $4.6\times3.5$ \\
  Line width (FWHM; km s$^{-1}$) & 420 & 390 & 420 & 450 \\
  Total flux (Jy km s$^{-1}$) & 168.3 & 14.1 & 20.9 & 3.7 \\
  NMA/IRAM 30 m flux ratio\footnotemark[$*$] & 0.71 & 0.67 & --- & 0.58 \\
  Luminosity (K km s$^{-1}$ pc$^2$) & $4.0\times10^9$ & $5.7\times10^8$ & $8.3\times10^8$ & $9.7\times10^7$ \\
  Ratio to $^{12}$CO (total)\footnotemark[$\dagger$] & --- & 0.14 & 0.21 & 0.024 \\
  Ratio to $^{12}$CO ($<$\timeform{4''.6}$\times$\timeform{3''.5})\footnotemark[$\dagger$] & --- & 0.25 & 0.37 & 0.047 \\
  Ratio to HCN (total)\footnotemark[$\dagger$] & 7.1 & --- & 1.5 & 0.17 \\
  Continuum flux (mJy) & --- & 16.6 & 16.6 & 10.8 \\ \hline
\multicolumn{5}{@{}l@{}}{\hbox to 0pt{\parbox{180mm}{\footnotesize
\par\noindent
\footnotemark[$*$] References for the IRAM 30m observations; $^{12}$CO and HCN: \citet{Solomon1992}, $^{13}$CO: Casoli, Dupraz, \& Combes (1992).
\par\noindent 
\footnotemark[$\dagger$] All ratios are in brightness temperature scale.
}\hss}}
\end{tabular}
\end{center}
\end{table}


\clearpage

\clearpage

\clearpage

\clearpage

\clearpage

\clearpage

\clearpage

\clearpage

\end{document}